\documentclass[12pt]{article}
\usepackage{graphicx,psfrag,color}       
\usepackage{multirow,rotating}       
\def\-{~-~}
\def\+{~+~}
\def\={~=~}
\def\eq{~\equiv~}
\def\beq{\begin{equation}}
\def\eeq{\end{equation}}
\def\beqn{\begin{eqnarray}}
\def\eeqn{\end{eqnarray}}
\def\bb{\begin{eqnarray*}}
\def\ee{\end{eqnarray*}}
\newcommand{\calle}[1]{(\ref{#1})}
\newcommand{\lp}{\left(}
\newcommand{\lb}{\left\lbrack}

\newcommand{\rp}{\right)}
\newcommand{\rb}{\right\rbrack}

\newcommand{\CR}{\mathcal{R}}
\newcommand{\CS}{\mathcal{S}}

\begin{document}

\begin{center}
{\Large\bf ADDING RESISTANCES AND CAPACITANCES\\
           IN INTRODUCTORY ELECTRICITY}

\vspace{5mm}
     
{\bfseries C.J. Efthimiou}\footnote{costas@physics.ucf.edu}
 ~~~and~~~
{\bfseries R.A. Llewellyn}\footnote{ral@physics.ucf.edu}\\
     Department of Physics\\
     University of Central Florida\\
     Orlando, FL 32826
\end{center}

\vspace{2cm}

\begin{abstract}
  We propose a unified approach to addition of resistors and capacitors
  such that the formul\ae\ are always simply additive. 
  This approach has the advantage of being consistent with the intuition
  of the students.
  To demonstrate
  our point of view, we re-work some well-known end-of-the-chapter
  textbook problems
  and propose some additional new problems.
\end{abstract}

\section{Introduction}

All introductory physics textbooks, with or without calculus, cover the addition
of both resistances and capacitances in series and in parallel.
The formul\ae\ for adding resistances
\beqn
\label{eq:1}
   R &=& R_1 + R_2 + \dots~, \\
\label{eq:2}
  {1\over R} &=& {1\over R_1} + {1\over R_2} + \dots~,
\eeqn
and capacitances
\beqn
\label{eq:3}
  {1\over C} &=& {1\over C_1} + {1\over C_2} + \dots~,\\
\label{eq:4}
   C &=& C_1 + C_2 + \dots~, 
\eeqn
are well-known and well-studied in all the books.

In books with calculus there are often end-of-chapter problems in which
students must
find $R$ and $C$  using continuous versions of equations 
\calle{eq:1} and \calle{eq:4} \cite{HRW,Serway,Tipler,WP,YF}. However, 
we have found \textit{none}  which includes problems that make
use of continuous versions of equations \calle{eq:2} and \calle{eq:3}
\cite{HRW,Hecht,Nolan,Serway,Tipler,WP,YF}.
Students who can understand and solve the first class of problems 
should be able
to handle the second class of the problems, as well.
We feel that continuous problems that make use of all four equations
should be shown to the students in order to give them a 
global picture of how calculus is applied to physical problems.
Physics contains much more than mathematics. When integrating quantities
in physics,
the way we integrate them is motivated by the underlying physics.
Students often forget
the physical reasoning and they tend to add  (integrate) quantities
only in one way.

In this paper, we introduce an approach to solving continuous
versions of equations \calle{eq:2} and \calle{eq:3} that is as straightforward
and logical
for the students as solving continuous versions of equations \calle{eq:1} 
and \calle{eq:4}. We then present some problems in which the student must
decide which formula is the right one to use for integration. We  hope 
that this article will motivate teachers to explain to students the subtle
points between `straight integration' as taught in calculus and 
`physical integration' to find a 
physical quantity.

\section{Adding Resistances}

\begin{figure}[h]
\begin{center}
\psfrag{r}{$r$}
\psfrag{dr}{$dr$}
\psfrag{a}{$a$}
\psfrag{l}{$l$}
\includegraphics[width=10cm]{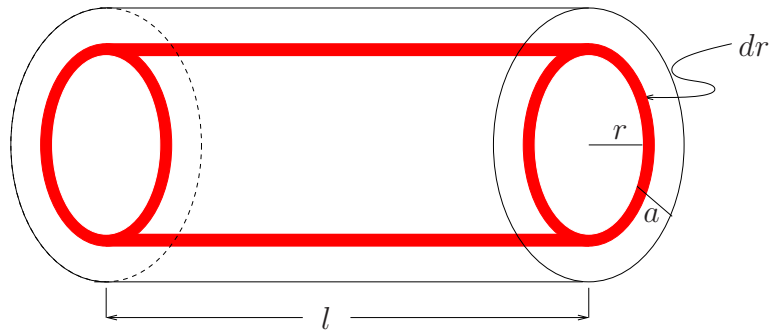}
\end{center}
\caption{The figure shows a  cylindrical wire of radius $a$. A potential 
         difference is
         applied between the bases of the cylinder and therefore electric
         current is running parallel to the axis of the cylinder.}
\label{fig:1}
\end{figure}

{\bfseries Problem} [Cylindrical Resistor]\\
The cylindrical resistor shown in figure \ref{fig:1} is made such that the
resistivity
$\rho$ is a function of the distance $r$ from the axis.
What is the total resistance $R$ of the resistor?

{\bfseries Towards a Solution}:

We divide the cylindrical resistor into infinitesimal resistors in the
form of cylindrical shells of thickness $dr$. One of these shells is seen
in red  in figure \ref{fig:1}.
If we apply equation \calle{eq:1} naively, we must write
$$
   R\=\int_{\rm cylinder} dR~.
$$
The infinitesimal resistance of the red
shell is given by
$$
    dR \= \rho(r)\, {\ell\over dA}~,
$$
where $dA=2\pi r dr$ is the area of the base of the infinitesimal shell.
Since $dr$ is small, $dR$ is huge, which is  absurd.  
Where is the error?

{\bfseries Discussion}:
    
When the current is flowing along the axis of the cylinder,
the infinitesimal resistors are not connected in series.
Therefore, the naive approach
$$
   R\=\int dR
$$
does not work since this formula assumes that the shells are
connected in series. Instead, all of the infinitesimal cylindrical shells
of width $dr$  are connected at the same end points and, therefore,
have the same applied potential. In other words, the shells
are connected in parallel and it is the inverse resistance that
is importnat, not $R$ itself. Specifically
$$
   {1\over R}\= \int_{\rm cylinder} d\lp{1\over R}\rp~.
$$
Students may feel unconfortable with this equation as at the begining since
it may seem `contradictory' to their calculus knowledge; therefore,
some discussion
may be helpful. 

Equation \calle{eq:1} states that when resistors are connected in
series, they make it harder for the current to go through. Their resistances
add to give the total restistance.
However, when resistors are connected in parallel, many `paths' are available
simultaneously; the current is flowing easily and `resistance' ---which is
a measure of flow difficulty--- is not a good quantity to use.  
Maybe an analogy from everyday life is useful here. Paying tolls at 
toll booths is in direct analogy. When only a single booth is
available, then all traffic has to go through that lane and no matter
how dense the traffic is, there will be a relative delay. The traffic
encounters some `resistance' in the flow. However, when multiple booths
are open, the drivers choose to go through the lanes that are free at the time
of their approach to the booths
and thus the delays encountered are minimal. In this case, the
`availability' of booths is a better quantity to be used to
descibe what is happening instead of the `resistance' at the booths.
Ultimately, the two quantities are related, but intuitely it is more
satisfying to use one over the over depending on the situation.
In direct analogy, for resistors connected in parallel,
the relevant quantity is not
$R$ any longer, but $S$, where
$$
     S \= {1\over R} \= \sigma\, {A\over\ell}~,
$$
and $\sigma=1/\rho$ is the conductivity.
$S$ is called the {\bfseries conductance} of the resistor. 
When resistors are connected in parallel,
they make it easier for the current to go through. Their conductances
add to give the total conductance:
$$
    S\= S_1 + S_1 + \cdots~.
$$
Thus, the conductance  follows the usual addition 
$$
   S \= \int dS
$$
when infitesimal resistors are connected in parallel.

We are now in position to compute the answer to the posed problem
in a way that is consistent with the intuition of the students.

{\bfseries Solution}:

Since the cylindrical shells are connected in parallel, conductance is
the additive quantity. For the infinitesimal shell
$$
  dS\=\sigma(r) \, {2\pi r dr\over\ell}~.
$$
Therefore
\bb
  S\=\int_{cylinder} dS \=  {2\pi\over\ell}\, \int_0^a  \sigma(r)\, r dr~.
\ee
For example, if $\sigma(r)\=\sigma_0 \,{a\over r}$, then
\bb
   S\= 2\sigma_0 \, {\pi a^2\over\ell} ~,
\ee
where $\sigma_0=1/\rho_0$.
The resistance is therefore 
\bb
    R\= {\rho_0\over2} \, {\ell\over \pi a^2}~.
\ee


{\bfseries Problem} [Truncated-Cone Resistor]\\
A resistor is made from a truncated cone of material with uniform
resistivity 
$\rho$. 
What is the total resistance $R$ of the resistor when the potential 
difference is applied between the two bases of the cone?

\begin{figure}[h]
\begin{center}
\psfrag{h}{$h$}
\psfrag{b}{$b$}
\psfrag{c}{$c$}
\psfrag{z}{$z$}
\psfrag{dz}{$dz$}
\psfrag{r}{$r$}
\includegraphics[width=7cm]{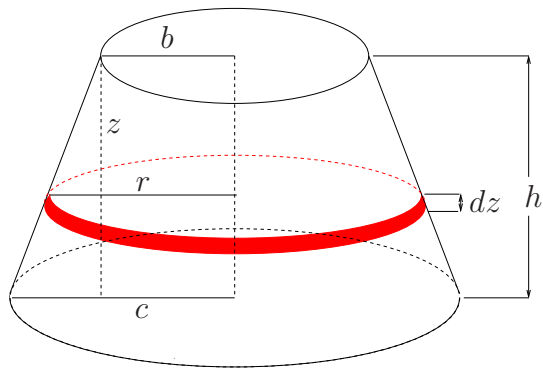}
\end{center}
\caption{A truncated cone which has been sliced in infitesimal cylinders of 
         height $dz$.}
\label{fig:cone}
\end{figure}

{\bfseries Solution}:

This is a well-known problem found in many of the introductory
physics textbooks \cite{HRW,Serway,Tipler,WP,YF}. 
We can partition the cone into infitesimal cylindrical resistors of
length $dz$. One representative resistor at distance $z$ from the top base
is seen in  figure \ref{fig:cone}. 
The area of the resistor is $A=\pi r^2$ and therefore its 
infinitesimal resistance is given by 
$$
   dR \= \rho\, {dz\over \pi r^2}~.
$$
From the figure we can see that
$$
  {z\over h}\= {r-b\over c-b} ~\Rightarrow~  dz\= {h\over c-b} dr ~.
$$

The infinitesimal resistors are connected in series and therefore
\beq
   R\=\int_{\rm cone} dR \= \rho\, {h\over\pi (c-b)} \, \int_b^c {dr\over r^2}
    \= \rho\,{h\over\pi bc} ~.
\label{eq:Rcone}
\eeq

{\bfseries Comment}:
{\footnotesize However, this solution, which is common in textbooks 
\cite{HRW,Serway,Tipler,WP,YF}, tacitly assumes that
the disks used in the partition of the truncated cone are equipotential
surfaces. This is of course not true, as can be seen quite easily.
If they were  equipotential surfaces, then the electric field lines
would be straight lines, parallel to the axis of the cone. However,
this cannot be the case as, close to the lateral surface of the cone,
it would mean that the current goes through the 
lateral surface and does not remain inside the resistor. Therefore,
the disks are not equipotential surfaces. One way out of this 
subtlety is to assume that
the disks are approximate equipotential surfaces as suggested in 
\cite{WP}. This is the attitude we adopt in this article as
our intention is not to discuss the validity of the partitions
used in each problem, but to emphasize the unified description
of resistances and capacitances as additive quantities. Similar
questions can be raised and studied in the majority of the problems 
mentioned in the present manuscript. A reader with serious interests in
electricity is referred to the article of  of Romano and Price
\cite{Romano} where the conical resistor is studied. Once that
article is understood, the reader can attempt to generalize
it to the rest of the problems of our article.}

\section{Adding Capacitances}

A similar discussion may be given for capacitors.
When capacitors are connected in parrallel, capacitance is the
the relative additive quantity:
$$
   C\=C_1+C_2+\cdots~.
$$
For a  parallel-plate capacitor of area $A$ and distance $d$ between the plates
$$
   C\=\varepsilon_0\,{A\over d}~.
$$
When the capacitor is filled with a uniform dielectric of  dielectric constant
$\kappa$ then
$$
   C\=\varepsilon_0\,\kappa\, {A\over d}~.
$$
However, when capacitors are connected in series, the inverse capacitance
$$
   D\= {1\over C}~.
$$
is the additive quantity. We may call it the {\bfseries incapacitance}.
For a  parallel-plate capacitor
$$
   D\= {1\over\varepsilon_0\kappa}\, {d\over A}~.
$$
In other words, when capacitors are connected in series 
$$
   D\= D_1+D_2+\cdots~.
$$
 
\begin{figure}[htb!]
\begin{center}
\psfrag{}{$$}
\includegraphics[width=10cm]{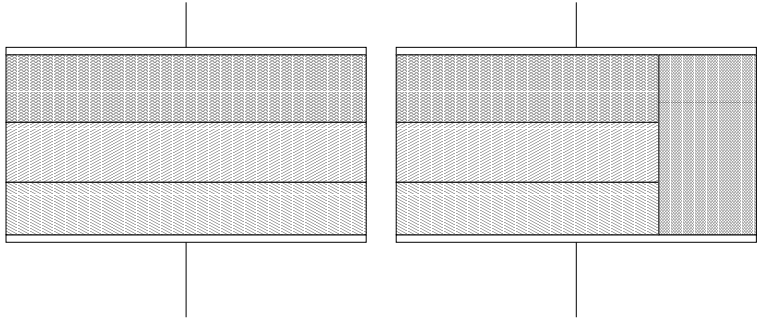}
\end{center}
\caption{Two parallel-plate capacitors which are filled with uniform
         dielectrics of different dielectric constants.}
\label{fig:dielectrics}
\end{figure}

Problems like this  are encountered when we fill a capacitor with
a dielectric for which the dielectric constact is a function of 
the distance from the plates of the capacitor.
Students are familiar with such problems for a parallel-plate capacitor
in the discrete case. For example, problems asking students to compute the
total capacitance in cases as those shown in figure \ref{fig:dielectrics}
are found in several textbooks \cite{Serway,Tipler,YF}. However, continuous
problems are not found in any textbook
\cite{HRW,Hecht,Nolan,Serway,Tipler,WP,YF}.

\begin{figure}[htb!]
\begin{center}
\psfrag{r}{$r$}
\psfrag{dr}{$dr$}
\psfrag{a}{$a$}
\psfrag{b}{$b$}
\psfrag{h}{$h$}
\includegraphics[height=10cm]{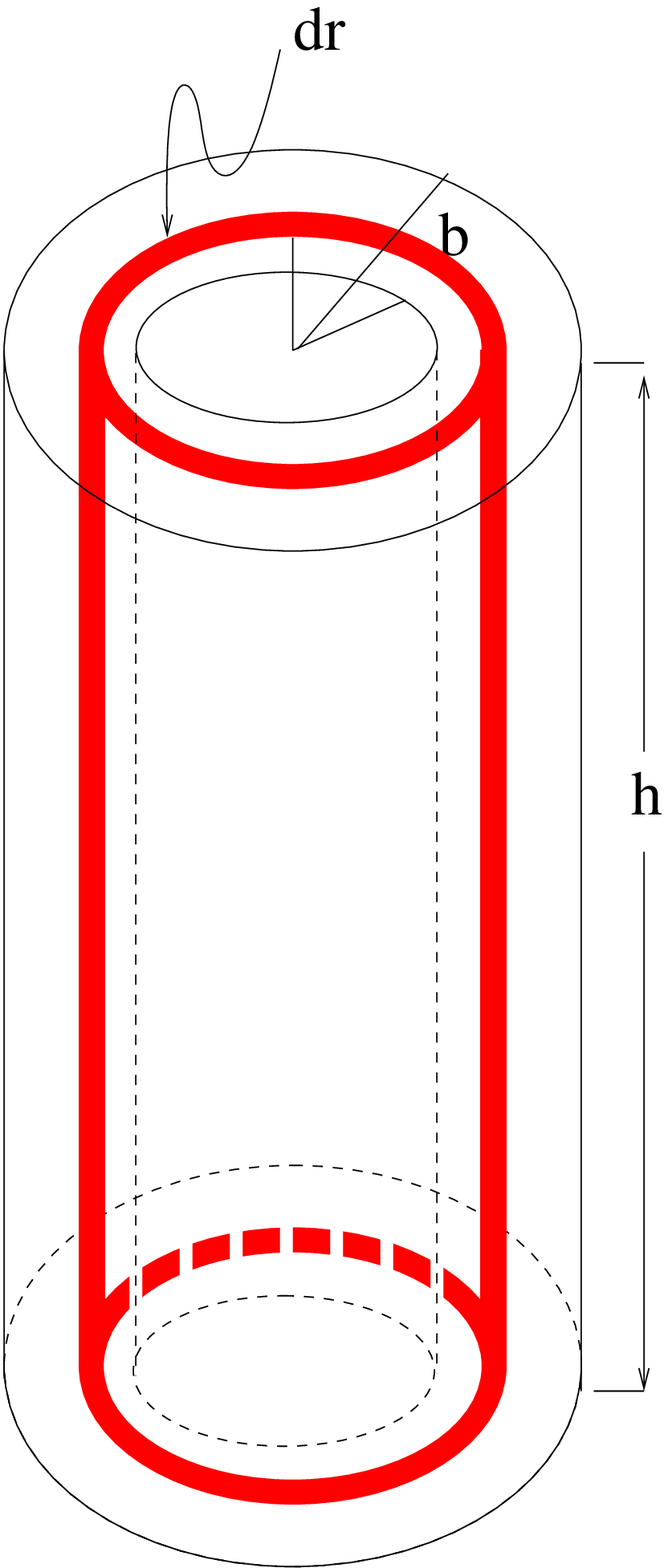}
\hspace{20mm}
\psfrag{r}{$r$}
\psfrag{dr}{$dr$}
\psfrag{a}{$a$}
\psfrag{b}{$b$}
\psfrag{h}{$h$}
\psfrag{dz}{$dz$}
\includegraphics[height=9cm]{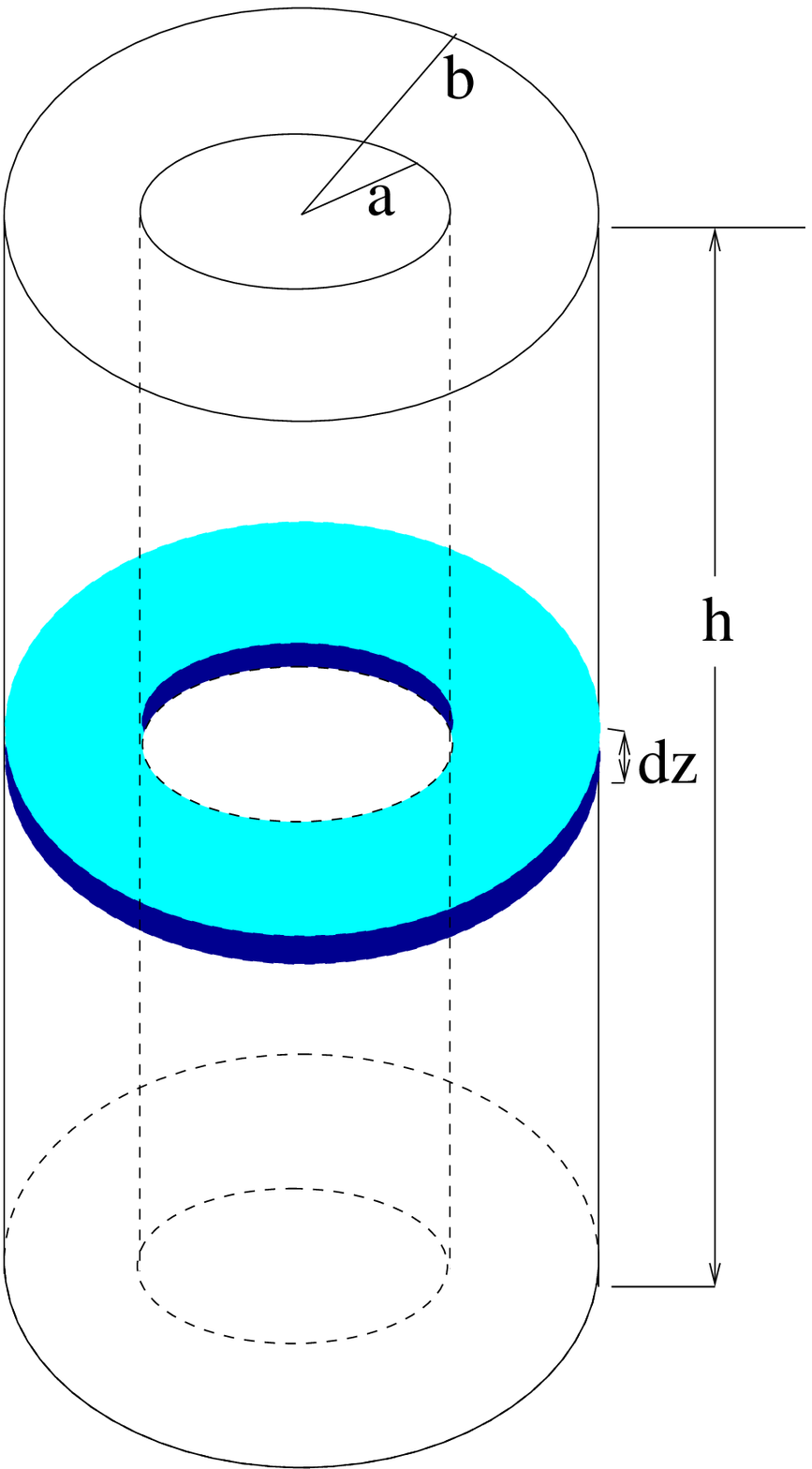}
\end{center}
\caption{A cylindrical capacitor with radii $a$ and $b$ and height $h$.
         In the left picture, we have sliced it in infinitesimal cylindrical
         shells, while in the right picture we have sliced it in infinitesimal
         annuli.}
\label{fig:hollowcylinder}
\end{figure}

We can easily construct new problems or re-work old problem using this
idea. For example, the well-known formula for the capacitance
of a cylindrical capacitor
can be found this way.
As shown in the left side of figure \ref{fig:hollowcylinder},
the capacitor is partitioned into small
cylindrical capacitors for which
the distance between the plates is $dr$. For such small capacitors,
the formula of a parallel-plate capacitor is valid. We notice though that
all infintesimal capacitors are connected in series. Therefore
$$
    dD ~=~ {1\over\varepsilon_0}\, {dr\over 2\pi r h}~.
$$
and
$$
    D ~=~ \int_{\rm cylinder} dD
      ~=~ {1\over2\pi\varepsilon_0h}\, \int_a^b{dr\over r}
      ~=~ {1\over2\pi\varepsilon_0h}\, \ln{a\over b}~.
$$
The total capacitance is then
$$
    C ~=~ {1\over D} ~=~ {2\pi\varepsilon_0 h\over\ln{b\over a}}~.
$$

{\bfseries Comment}:
\footnotesize
One might be tempted to
partition the cylindrical capacitor into infinitesimal
capacitors 
as seen in the figure to the left (blue section). Such capacitors
look simpler than the  
infitesimal cylindrical shell we used above. Furthermore, they
are connected in
parallel (notice that each capacitor is carrying an infinitesimal charge $dQ$
and $\int_{\rm cylinder} dQ=Q$) and  therefore it is enough to deal with
capacitance,
$C=\int_{\rm cylinder} dC$, and not incapacitance $D$.

However, with a minute's reflection the reader will
see that in order to use the parallel-plate capacitor formula in the
infinitesimal case, the distance between the plates must be infinitesimal
which indicates that the infinitesimal capacitors must
be connected in series.
In the proposed (blue) slicing, the distance between the plates
of the infinitesimal capacitor is finite, namely $b-a$. The infinitesimal
capacitor is still a cylindrical capacitor of infitesimal height and therefore
its capacitance should be expressed in a form that is  not known before the
problem is solved. In other words,
\beq
   dC ~=~ {2\pi\varepsilon_0\over\ln(b/a)} \, dz
\label{eq:unknown}
\eeq
from which
\beq
   C ~=~ {2\pi\varepsilon_0\over\ln(b/a)} \, \int_0^h dz
     ~=~ {2\pi\varepsilon_0\over\ln(b/a)} \,h~.
\label{eq:known}
\eeq
But the expression \calle{eq:unknown} is unkown until the
result \calle{eq:known} is found.
\normalsize

\vspace{5mm}
{\bfseries Problem} [Truncated-cone Capacitor]

A capacitor is made of two circular disks of radii $b$ and $c$ respectively
placed at a distance $h$ such that the line that joins their centers is
perpendicular to the disks. Find the capacitance of this arrangement
(Seen in figure \ref{fig:cone}).

{\bfseries Solution}: We partition the capacitor into infinitesimal
parallel-plate capacitors of distance $dz$ and plate area $A=\pi r^2$
exactly as seen in figure \ref{fig:cone}. These infitesimal
capacitors are connected in series and therefore the incapacitance
is the relevant additive quantity:
$$
   dD \= {1\over\varepsilon_0}\,{dz\over \pi r^2}~.
$$
Notice that the computation is identical to that of $R$ with final result:
\beq
  D \= {1\over\varepsilon_0}\, {h\over\pi bc}
  ~\Rightarrow~ C\= \varepsilon_0 \, {\pi bc\over h}~.
\label{eq:trunccone}
\eeq
When $b=c$, we recover the result of the parallel-plate capacitor.

\section{Conclusions}

In this paper, we have tried to argue that when the right variables
are used then the law of addition for capacitances and resistances
is always additive. Table \ref{table:1} summarizes our
main formul\ae.  This is in agreement with the intuition
of students when they solve continuous problems on the subject
who like to add quantities in a simple way.

\newcommand{\paral}{{\bfseries\textcolor{red}{parallel}}}
\newcommand{\series}{{\bfseries\textcolor{red}{series}}}
\newcommand{\resistors}{{\bfseries\textcolor{blue}{resistors}}}
\newcommand{\capacitors}{{\bfseries\textcolor{blue}{capacitors}}}
\newcommand{\inductors}{{\bfseries\textcolor{blue}{inductors}}}
\newcommand{\springs}{{\bfseries\textcolor{blue}{springs}}}
\newcommand{\thermal}{{\bfseries\textcolor{blue}{thermal}}}
\newcommand{\conductors}{{\bfseries\textcolor{blue}{conductors}}}
\begin{table}[htb!]
\begin{center}
\begin{tabular}{|c|c|c|} \hline
         & \resistors & \capacitors \\ \hline
         & & \\ 
   \series  & $R=\sum\limits_i R_i$   &  $D=\sum\limits_i D_i$ \\ 
         & & \\ \hline 
         & & \\ 
   \paral  & $S=\sum\limits_i S_i$   &  $C=\sum\limits_i C_i$ \\ 
         & & \\ \hline 
\end{tabular}
\end{center}
\caption{When viewed in the right physical quantities, addition
         of resistors and capacitors is always simple. In the above table,
         $R$ is the resistance, $S$ is the conductance, $C$ is the capacitance,
         and $D$ the incapacitance of a circuit element.}
\label{table:1}
\end{table}

We must point out that our discusssion by no means is restricted to
capacitances and resistances only. Similar addition  laws are encountered
in many other areas of physics. For example, when connecting springs
in parallel the total stiffness constant is given by the sum
of the individual stiffness constants:
$$
    k \= k_1 + k_2 + \cdots ~.
$$
When the springs are connected in series, then
$$
    {1\over k} \= {1\over k_1} + {1\over k_2} + \cdots ~,
$$
pointing out that in this case, not the stiffness constant but
the {\bfseries elasticity constant} 
$$
   \ell\={1\over k}
$$
is the relevant additive constant. Our discussion can thus be repeated
verbatim in all similar cases. Table \ref{table:2} lists the most
common cases found in introductory physics.

\begin{table}[htb!]
\begin{center}
\begin{tabular}{|c|c|c|c|c|c|} \hline
         & & & & & \\ 
         & \resistors & \capacitors & \inductors & \springs & \thermal \\ 
         & & & & & \conductors \\ 
         & & & & & \\ \hline
         & & & & & \\ 
   \series  & $\matrix{\mbox{resistance}\cr R\cr}$ 
            & $\matrix{\mbox{incapacitance}\cr D\cr}$
            & $\matrix{\mbox{inductance}\cr L\cr}$
            & $\matrix{\mbox{elasticity}\cr \ell\cr}$ 
            & $\matrix{\mbox{thermal}\cr \mbox{resistance}\cr\CR\cr}$ \\
         & & & & & \\ \hline 
         & & & & & \\ 
   \paral  & $\matrix{\mbox{conductance}\cr S\cr}$
            & $\matrix{\mbox{capacitance}\cr C\cr}$
            & $\matrix{\mbox{deductance}\cr K\cr}$
            & $\matrix{\mbox{stiffness}\cr k\cr}$ 
            & $\matrix{\mbox{thermal}\cr\mbox{conductance}\cr\CS\cr}$ \\
         & & & & & \\ \hline 
\end{tabular}
\end{center}
\caption{This table summarizes the additive physical quantities in the most
         common cases encountered in introductory physics. The quantities 
         that are not usually defined in the introductory books are
         the conductance $S=1/R$, the incapacitance $D=1/C$, the deductance
         $K=1/L$, the elasticity constant $\ell=1/k$, and the the thermal
         conductance
         $\CS=1/\CR$.}
\label{table:2}
\end{table}

\newpage \ \newpage
\appendix

\section{Suggested Problems}

We end our article with some suggested problems which the reader may
wish to solve. 

\begin{enumerate}

\item
Re-derive the well-known expression for the capacitance of a spherical capacitor
$$
    C \= 4\pi\varepsilon_0 {ab\over b-a}~,
$$
(where $a, b$ are the radii of the spheres with $b>a$)
by partitioning it into
infinitesimal capacitors.

\item
Show that the capacitance of a cylindrical capacitor which is filled with
a dielectric having dielectric constant $\kappa(r)=c r^n$, where $r$
is the distance from the axis and $c$, $n\ne 0$ are constants, is given
by
$$
    C \= 2\pi\varepsilon_0 \, hcn\, { a^n b^n\over b^n-a^n}~.
$$

\item
Show that the capacitance of a cylindrical capacitor which is filled with
a dielectric having dielectric constant $\kappa(z)=c z^n$, where $z$
is the distance from the basis and $c$, $n\ge 0$ are constants is given
by
$$
    C \= 2\pi\varepsilon_0 \, {ch^{n+1}\over(n+1)\,\ln(b/a)}~.
$$

\item
Show that the capacitance of a spherical capacitor which is filled with
a dielectric having dielectric constant $\kappa(r)=c r^n$, where $r$
is the distance from the center and $c, n$ are constants is given
by
$$
    C \= 4\pi\varepsilon_0 \, {c\over\ln(b/a)}
$$
for $n=-1$ and
$$
    C \= 4\pi\varepsilon_0 \, c\,(n+1)\, 
         {a^{n+1} b^{n+1}\over b^{n+1}-a^{n+1} }
$$
for $n\ne-1$.

\begin{figure}[h]
\begin{center}
\psfrag{r}{$r$}
\psfrag{dr}{$dr$}
\psfrag{a}{$a$}
\includegraphics[width=7cm]{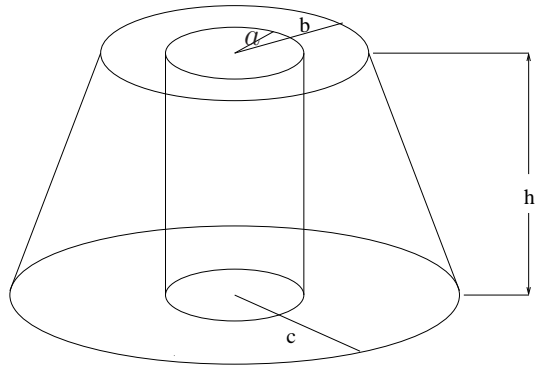}
\end{center}
\caption{A hollow truncated cone}.
\label{fig:hollowcone}
\end{figure}

\item 
(a) Two metallic flat annuli are placed such that they form a
capacitor with the shape of a hollow truncated cone as seen
in figure \ref{fig:hollowcone}. Partition the capacitor in 
infitesimal capacitors and show that the capacitance 
is given by
$$
    C \= 2\pi\varepsilon_0\, {h\over a(c-b)}\,
         \lb \ln{c-a\over c+a}-\ln{b-a\over b+a}\rb~.
$$
Show that this result reduces to that of a cylindrical capacitor for
$c=b$. Also, show that it agrees with \calle{eq:trunccone} when
$a=0$.
   
(b) Now, fill the two bases with disks of radius $a$ and argue that
the capacitance of the hollow truncated cone equals that of the
truncated cone minus the capacitance of the parallel-plate
capacitor that we have removed (superposition principle).
This means that the capacitance of
the hollow truncated cone should equal to
$$
   C\= \pi\varepsilon_0\, {bc-a^2\over h}~.
$$
How is it possible that this result does not agree with that of part
(a)?

\item 
A capacitor with the shape of a hollow truncated cone is now formed
from two `cylindrical' shells. Show that the capacitance in this case is
$$
   C\= 2\pi\varepsilon_0\,{ah\over c-b}\, 
      \lb {\rm li}\lp{c\over a}\rp- {\rm li}\lp{b\over a}\rp \rb ~,
$$
where $\mbox{li}$ is the logarithmic function\cite{AS}
$$
   {\rm li}(X) \eq \int_0^X {dx\over\ln x}~,~~~~~X>1~.
$$ 

\item
(a) A conductor has the shape seen in figure \ref{fig:hollowcone}.
Show that the resistance
when the voltage is applied between the upper and lower
bases is given by
$$
  R\=\rho\,{h\over 2\pi a(c-b)}\,
     \lb \ln{c-a\over c+a}-\ln{b-a\over b+a}\rb~.
$$
Show that this result reduces to equation \calle{eq:Rcone}
for $a=0$. 

(b) Argue now that the resistance of the hollow truncated-conical
wire is the difference between the the resistance of the truncated-conical
wire and a cylindrical wire of radius $a$ (superposition principle).
This implies that 
$$
     R\=\rho\,{h\over bc-a^2}~.
$$
Explain why this does not agree with part (a).

\item
A conductor has the shape of a hollow cylinder as seen in figure
\ref{fig:hollowcylinder}.
Show that the resistance
when the voltage is applied between the inner and outer surfaces
is given by
$$
   R \={\rho\over 2\pi h}\, \ln{b\over a}~.
$$

\item
A conductor has the shape seen in figure \ref{fig:hollowcone}.
Show that the resistance
when the voltage is applied between the inner and outer surfaces
is given by
$$
   R \={\rho\over 2\pi h a}\, {c-b\over{\rm li}(c/a)-{\rm li}(b/a)}~.
$$
Show that, for $c=b$, this result agrees with that of
the previous problem.

\item
A conductor has the shape of a truncated wedge as seen in figure
\ref{fig:wedge}.
Show that the resistance of the conductor
when the voltage is applied between the left and right faces
is 
$$
   R\= {\rho\over a}\, {\ell \,\ln(c/b)\over c-b}~,
$$
while the resistance when the voltage is applied between the top and
bottom faces is
$$
   R\= {\rho\over a}\, {c-b\over\ell\,\ln(c/b)}~.
$$

\begin{figure}[h]
\begin{center}
\psfrag{x}{$x$}
\psfrag{y}{$y$}
\psfrag{z}{$z$}
\psfrag{a}{$a$}
\psfrag{b}{$b$}
\psfrag{c}{$c$}
\psfrag{l}{$\ell$}
\includegraphics[width=7cm]{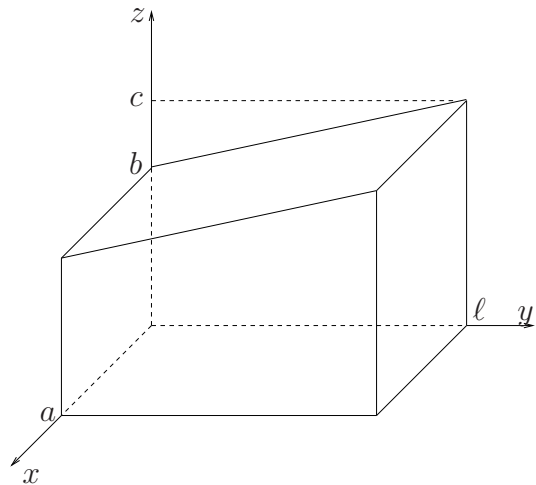}
\end{center}
\caption{A conductor with the shape of a truncated wedge.}
\label{fig:wedge}
\end{figure}
\end{enumerate}


\end{document}